\documentclass[prl,aip,amssymb,twocolumn,showpacs,floatfix]{revtex4}

\usepackage{graphicx}
\usepackage{dcolumn}
\usepackage{bm}

\begin{document}

\preprint{}

\title{Bias-dependent electron spin lifetimes in \emph{n}-GaAs and the role of donor impact ionization}

\author{M.~Furis$^{1}$, D. L. Smith$^{2}$, J. L. Reno$^{3}$, and S.~A.~Crooker$^{1}$\email{crooker@lanl.gov}}

\affiliation{$^{1}$National High Magnetic Field Laboratory, Los Alamos, NM 87545}

\affiliation{$^{2}$Theoretical Division, Los Alamos National
Laboratory, Los Alamos, NM 87545}

\affiliation{$^{3}$Sandia National Laboratories, Department 1123, MS 1315, Albuquerque, New Mexico 87185-1315}

\date{\today}

\begin{abstract}

In bulk \emph{n}-GaAs epilayers doped near the metal-insulator transition, we study the evolution
of electron spin lifetime $\tau_s$ as a function of applied lateral electrical bias $E_x$. $\tau_s$
is measured via the Hanle effect using magneto-optical Kerr rotation. At low temperatures ($T<10$K,
where electrons are partially localized and $\tau_s > 100$ ns at zero bias), a marked collapse of
$\tau_s$ is observed when $E_x$ exceeds the donor impact ionization threshold at $\sim$10 V/cm. A
steep increase in the concentration of warm \emph{de}localized electrons -- subject to
Dyakonov-Perel spin relaxation -- accounts for the rapid collapse of $\tau_s$, and strongly
influences electron spin transport in this regime.

\end{abstract}


\maketitle

The discovery of very long electron spin lifetimes, $\tau_s$, in GaAs has helped motivate
considerable interest in semiconductor-based spintronic devices \cite{ss}. Studies of bulk
\emph{n}-GaAs have established that, in the absence of any applied electric field, long spin
lifetimes in the range $\tau_s$=100-300 ns are associated with cryogenic temperatures and electron
doping in the range of the metal-insulator transition ($n_{MIT} \simeq2\times10^{16}$ cm$^{-3}$)
\cite{kikkawa,dzhioev1,dzhioev2,colton}. Several prototype devices operating at low temperatures
have therefore incorporated \emph{n}-GaAs doped in this range
\cite{hagele,spinhall,stephens,li,jiang,crooker1,lou}. This is, however, precisely the doping and
temperature regime where carrier transport becomes highly nonlinear due to electron localization
(freeze-out) and subsequent impact ionization of donor-bound
electrons at quite modest applied electric fields of order 10 V/cm \cite {reynolds, oliver}. Since
future spin-transport devices may employ electrical biases on (at least) this order, a
bias-dependent study of $\tau_s$ through the threshold of donor impact ionization is important for device design and for the characterization of spin transport in \emph{n}-type semiconductors in general.

Here we measure $\tau_s$ as a function of temperature and applied electrical bias $E_x$ in bulk
epilayers of \emph{n}-GaAs doped near the metal-insulator transition. $\tau_s$ is obtained from
Hanle depolarization measurements using the magneto-optical Kerr effect.  At low temperatures,
$\tau_s$ drops dramatically at the threshold of donor impact ionization at $E_x \simeq 10$ V/cm.
At this threshold, which coincides with marked nonlinearities in the samples' current-voltage
(\emph{I-V}) characteristics, the electron ensemble becomes largely delocalized (and warm) and the
Dyakonov-Perel spin relaxation mechanism dominates.

Three silicon-doped bulk \emph{n}-GaAs epilayers were grown by molecular beam epitaxy on
(001)-oriented GaAs substrates. The dopings and thicknesses are: $1 \times 10^{16}$ cm$^{-3}$ (1
$\mu$m thick), $5 \times 10^{16}$ cm$^{-3}$ (1 $\mu$m thick), and $4 \times 10^{15}$ cm$^{-3}$ (15
$\mu$m thick). These samples allow study of the bias-dependent spin lifetime both above and below
the metal-insulator transition (MIT). The inset of Figure 1(a) shows the experimental geometry. The
samples are mounted, strain-free, on the cold finger of an optical cryostat. Ohmic contacts of
annealed indium or AuGeNi permit application of an in-plane electric field, $E_x$. Spin-polarized
electrons are optically injected by a 1.58 eV cw diode laser defocused to a large spot ($>$500 $\mu$m spot; the pump spot size must exceed the spin drift and diffusion lengths for accurate $\tau_s$
studies). The pump laser polarization is modulated from left- to right-circular
(injecting spins along $\pm \hat{z}$) at 50 kHz to enable lock-in detection and to minimize the
buildup of nuclear spin polarization via hyperfine interactions \cite{dzhioev2,oo}. The average
$\hat{z}$-component of electron spin in the epilayer ($S_z$) is detected via the polar
magneto-optical Kerr effect using a narrowband cw probe laser focused to a 4 $\mu$m spot and tuned
to a photon energy several meV below the GaAs band-edge \cite{crooker2}. The polarization (Kerr) rotation
$\theta_K$ imparted on the reflected probe laser is proportional to $S_z$ (for both localized or
free electron spins). Helmholtz coils provide an in-plane magnetic field ($B_y$) along $\hat{y}$,
and additional coils along $\hat{x}$ and $\hat{z}$ null the geomagnetic field at the sample.

We measure $\tau_s$ via the Hanle effect; that is, by measuring the depolarization of $S_z$ by the
transverse magnetic field $B_y$, analogous to established methods based on the
Hanle depolarization of photoluminescence \cite{dzhioev1,dzhioev2,oo}. In this geometry, the Hanle
curves exhibit Lorentzian lineshapes, $S_z(B_y)=S_0/[1+ (g_e \mu_B B_y \tau_s/\hbar)^2]$, with
half-widths $B_{1/2} = \hbar/g_e \mu_B \tau_s$.  Here, $g_e$ is the electron g-factor, $\mu_{B}$ is
the Bohr magneton, and $S_0$ is the overall amplitude.  Assuming $g_e$=-0.44 in bulk GaAs, a
half-width $B_{1/2}$=1 Gauss corresponds to $\tau_s$=258 ns.  The inset of Fig. 1(b) shows raw
Hanle data from the $1 \times 10^{16}$ cm$^{-3}$ epilayer at 4~K in lateral electric fields
$E_x$=8, 11, and 14 V/cm. The half-widths $B_{1/2}$=1.4, 2.6, and 5.5 G correspond to $\tau_s$=184,
99, and 47 ns respectively.  We use very low excitation ($<$1
mW/cm$^2$): $B_{1/2}$ is independent of pump (and probe) laser power, and $S_0$ scales linearly
with pump and probe power.

\begin{figure}[tbp]
\includegraphics[width=.40\textwidth]{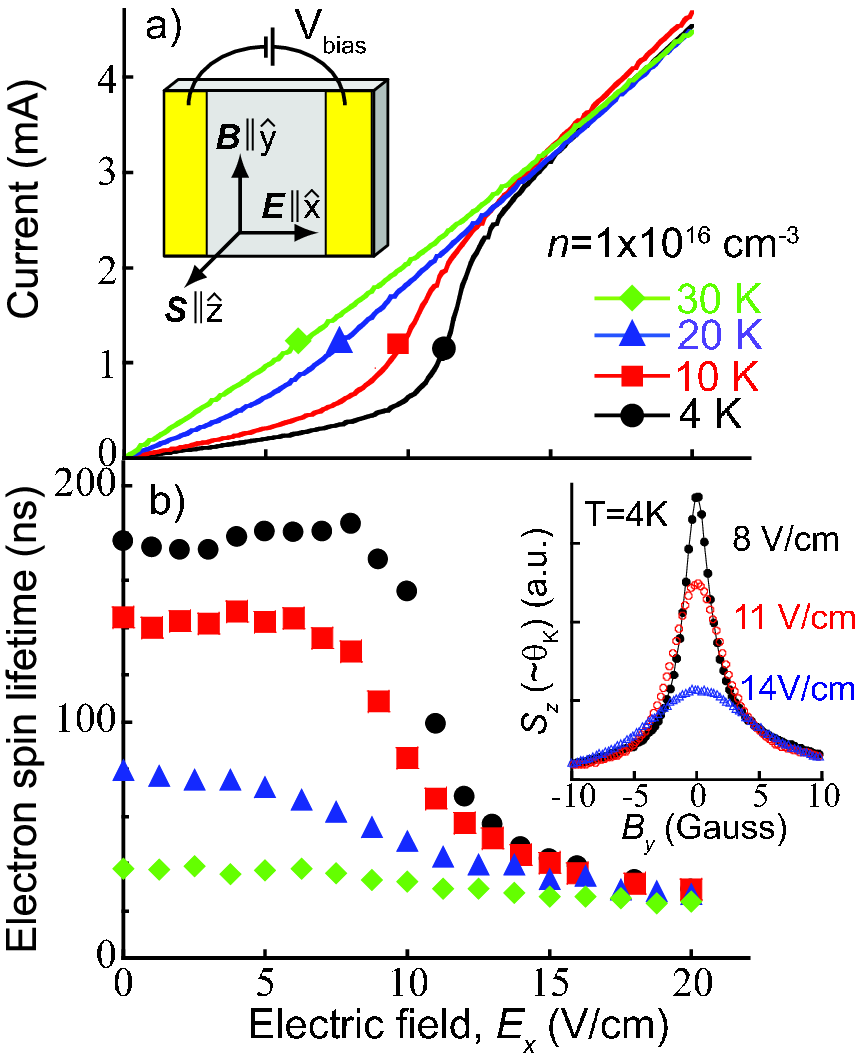}
\caption{a) Current vs. in-plane electric field $E_x$ in the $n$=$1\times 10^{16}$ cm$^{-3}$
\emph{n}-GaAs epilayer at $T$=30, 20, 10, 4~K. At low $T$, impact ionization of donors occurs at
$E_x \sim 11$ V/cm. Inset: The experimental geometry. b) The corresponding electron spin lifetime
$\tau_s$, measured via the Hanle effect. Inset: Hanle data ($S_z$ vs. $B_y$) at
4~K, in the regime of impact ionization.} \label{fig1}
\end{figure}

Figure 1(a) shows \emph{I-V} transport data from the $n$=$1 \times 10^{16}$ cm$^{-3}$ epilayer (doped
just below the MIT) at 30, 20, 10, and 4 K.  In the regime of small in-plane
electric fields ($E_x < 10$ V/cm) the resistance increases as the temperature falls, due to
localization (freeze-out) of electrons onto donor sites. The \emph{I-V} curves are approximately
linear in this regime; at the lowest temperatures most electrons are localized on the Si
donors and the small concentration of thermally-excited free electrons is independent of $E_x$. At
4~K, strong \emph{I-V} nonlinearities at $E_x \sim 11$ V/cm indicate the threshold of impact
ionization in this epilayer, in agreement with prior studies of \emph{n}-GaAs
\cite{reynolds,oliver}. At this threshold, sufficiently many free electrons possess the kinetic
energy required to impact-ionize the donor-bound electrons, resulting in a breakdown process (the donor binding energy is $\simeq$4
meV, or $\simeq$40~K). This threshold marks the transition between
an electron ensemble that is largely localized and cold (at low $E_x$), and an ensemble that is
largely free and warm -- of order 40 K -- at high $E_x$.

The corresponding average spin lifetime $\tau_s$ in this sample is shown in Fig. 1(b). At 4~K, $\tau_s$ is long ($\simeq$175 ns) and relatively independent of electrical bias for $E_x\lesssim10$
V/cm. In the narrow bias range between 10-14~V/cm, just beyond the ionization threshold, $\tau_s$ drops rapidly by a factor of three, and thereafter decreases more gradually. Similar (though less pronounced) trends are observed at 10 and 20~K. By 30~K, most donors are already thermally ionized at zero bias, so that electrons are predominantly free and the \emph{I-V} curves are nearly linear, and only gradual changes in $\tau_s$ are observed. The sudden collapse of $\tau_s$ at the lowest temperatures results directly from the impact ionization of cold donor-bound electrons into warm free electron states that are subject to efficient Dyakonov-Perel (DP) spin relaxation \cite{oo,dyakonov}. The DP mechanism originates in the spin-orbit splitting of finite-momentum states in the conduction band, and therefore dominates the spin relaxation of \emph{free} electrons in bulk \emph{n}-GaAs. Donor-bound electrons, being localized, are not subject to DP spin relaxation (rather, $\tau_s$ of donor-bound electrons is limited to a few hundred ns by hyperfine interactions with nuclear spins \cite{dzhioev2,oo}). Within the DP formalism for free electrons, $\tau_s$ decreases rapidly with electron energy $U$: $\tau_s \propto 1/U^{3} \tau_{p}(U)$, where $\tau_p$ is the momentum scattering time. For degenerate free electrons (Fermi energy $E_F \gg k_B T$), the electron energy $U\propto E_F$. At high temperatures where $k_B T \gg E_F$, $U\propto k_B T$. Regardless, an additional electrical bias $E_x$ increases $U$ by an average amount $m_e \mu^2 E_x^2/2$ in the simplest approximation, where $m_e$ and $\mu$ are the electron mass and energy-dependent mobility, respectively. As a result, the DP mechanism reduces $\tau_s$ with increasing $E_x$ due to the increased average electron energy.  For free electrons and a large electrical bias, a detailed calculation of this effect and its influence on $\tau_s$ was reported by Beck et al \cite{beck}.

The collapse of $\tau_s$ can therefore be associated with the sudden increase in average electron energy (\emph{i.e.}, electron temperature) resulting from donor impact ionization. Just beyond the threshold of impact ionization, most electrons are free and are necessarily at rather warm temperatures of order the donor binding energy ($\simeq$40 K). As such, $\tau_s$ is correspondingly short ($\simeq$30 ns, in agreement with past work \cite{kikkawa}).  Note that all $\tau_s(E_x)$ data converge for $E_x >$15 V/cm, supporting the notion of a common (and warm) electron temperature regardless of the zero-bias sample temperature.

\begin{figure}[tbp]
\includegraphics[width=.40\textwidth]{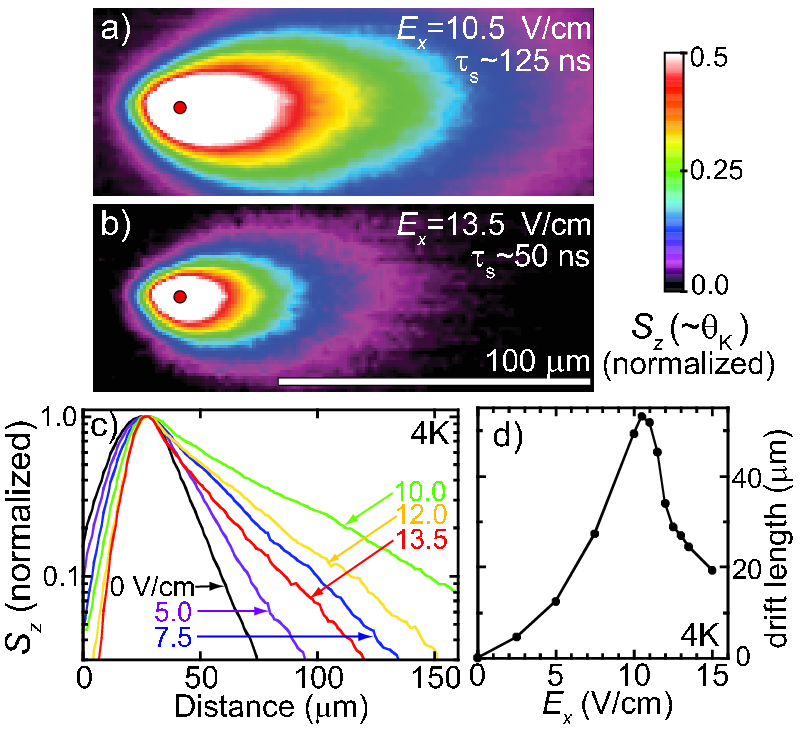}
\caption{a,b) 60$\times$160 $\mu$m images showing spin-polarized electron transport in the $n=1
\times 10^{16}$ cm$^{-3}$ \emph{n}-GaAs epilayer at 4~K. Spin-polarized electrons, optically
injected over a 4 $\mu$m spot (red dot), subsequently diffuse and drift in $E_x$. Images acquired
just before ($E_x$=10.5 V/cm) and after (13.5 V/cm) the onset of donor impact ionization. c)
Line-cuts through a series of images. d) The measured spin drift length vs. $E_x$.} \label{fig2}
\end{figure}

The collapse of $\tau_s$ at low temperatures naturally has a pronounced effect on spin
\emph{transport} in \emph{n}-GaAs. Figure 2 shows 60$\times$160 $\mu$m images of spin-polarized
electron flow at 4~K, just below and above the threshold of impact ionization.  Here, spins are
\emph{locally} injected by focusing the pump laser to a small 4 $\mu$m spot, and the probe laser is
raster-scanned in the \emph{x-y} sample plane (details can be found in Ref. \cite{crooker2}). The
images clearly show the effect of impact ionization on spin transport:  just prior to threshold at
$E_x$=10.5 V/cm, lateral drift of spins exceeds 120 $\mu$m, due in large part to their long spin
lifetime. A small increase of $E_x$ to 13.5 V/cm - just above threshold - results in a significant
decrease in the spin transport length owing to the collapse of $\tau_s$. Fig. 2c shows normalized
line-cuts through selected 4~K images. At $E_x$=0, diffusion drives spin transport over the spin
diffusion length $\sqrt{D \tau_s} \simeq 8$ $\mu$m. At large $E_x$, electron drift dominates spin
transport \cite{flatte}, approaching the characteristic lengthscale $\mu E_x \tau_s$. On the downstream side of the injection point, the observed $1/e$ spin decay
length, $L_d$, reaches a maximum at $E_x$=10.5 V/cm, while the upstream decay length $L_u$
shrinks with $E_x$, eventually reaching the 4 $\mu$m resolution limit.  Fig. 2d shows the measured
4~K spin drift length, $L_d-L_u=\mu E_x \tau_s$ \cite{flatte}, which is strongly peaked at the ionization threshold.

Figure 3 shows a similar correspondence between low-temperature transport and $\tau_s(E_x)$ measurements in the other \emph{n}-GaAs epilayers.  In the $n$=$5 \times 10^{16}$ cm$^{-3}$ epilayer, the \emph{I-V} curves show only slight indications of impact ionization even at 4~K, and $\tau_s(E_x)$ exhibits a more gradual decrease. This sample is doped above the MIT, and electrons are never completely localized. In contrast, transport in the $n$=$4 \times 10^{15}$ cm$^{-3}$ epilayer (doped well below the MIT) clearly shows electron localization and impact ionization at $E_x$=9.5 V/cm.  At low $T$  and $E_x$, the Lorentzian Hanle curves have surprisingly narrow half-widths ($B_{1/2}\lesssim 0.6$ G) that are independent of pump and probe laser power, field sweep rate (0.8 s/G to 40 s/G), and pump polarization modulation ($\pm \lambda/4$ to $\pm \lambda/20$). While these checks alone cannot definitively rule out the possible influence of finite nuclear spin polarization \cite{oo,paget}, a very rapid collapse of $\tau_s$ is nonetheless observed at the 4~K donor ionization threshold, in accord with the much sharper \emph{I-V} nonlinearities in this sample. In both samples, the $\tau_s(E_x)$ data converge above the donor ionization threshold regardless of lattice temperature, indicating DP spin relaxation of similarly warm electron ensembles.

\begin{figure}[tbp]
\includegraphics[width=.45\textwidth]{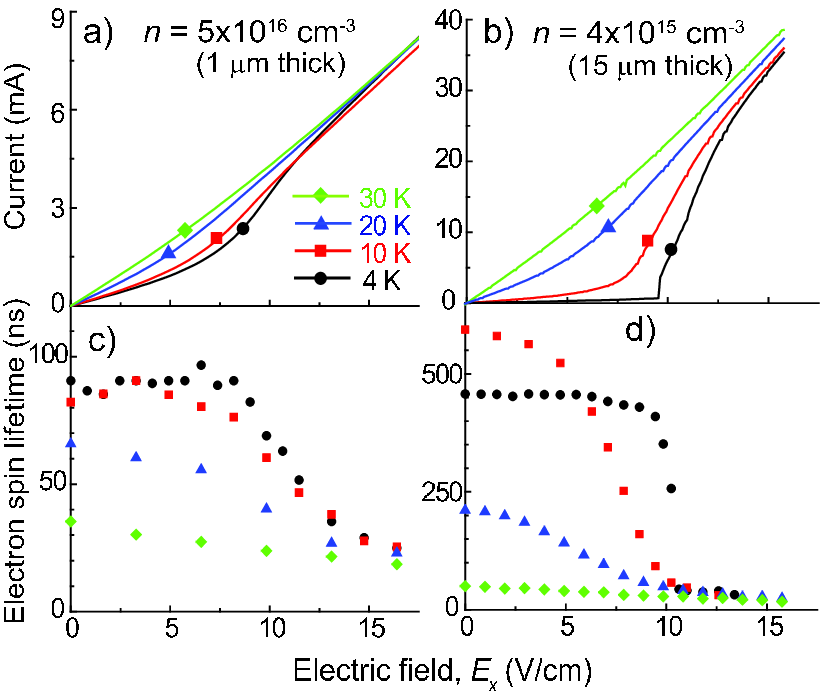}
\caption{a,b) Current vs. in-plane electric field $E_x$ in the $n=5 \times 10^{16}$ and $n=4
\times 10^{15}$ cm$^{-3}$ \emph{n}-GaAs epilayers. c,d) Corresponding electron spin lifetimes.}
\label{fig3}
\end{figure}

In conclusion, the long electron spin lifetimes in low-temperature \emph{n}-GaAs decrease markedly
at the onset of donor impact ionization at $E_x \sim $10 V/cm. This threshold marks the transition
from an ensemble of largely localized spins to an ensemble of primarily free - and warm - electron
spins subject to efficient Dyakonov-Perel spin relaxation.  This collapse, observed at relatively
low biases that are commonly exceeded in semiconductor devices, is important for prototype spin
transport devices which rely on the interplay between electron mobility and spin lifetime to
achieve desired spin transport lengths. We thank Paul Crowell for valuable discussions, and
acknowledge support from the Los Alamos LDRD program.


\end{document}